%% file: main_ICASSP_v3__camera-ready_.tex
\documentclass[conference,romanappendices]{IEEEtran}
\IEEEoverridecommandlockouts

\usepackage{algorithm,amsmath,amssymb,amsthm,bbm,cite,color,comment,enumitem,flushend,graphicx,microtype,setspace,url,algpseudocode,setspace,subcaption}
\usepackage[USenglish]{babel}
\usepackage{hyperref}
\usepackage[utf8]{inputenc}
\usepackage[T1]{fontenc}

\usepackage[labelsep=period,font=small]{caption}
\usepackage{tikz,pgfplots}
\usetikzlibrary{arrows, arrows.meta, backgrounds, calc, intersections, decorations.markings, decorations.pathreplacing, positioning, shapes}
\pgfplotsset{compat=1.17}

\input{./Definitions.tex}

\algnewcommand\algorithmicinitialize{\textbf{Initialize:}}
\algnewcommand\Initialize{\item[\algorithmicinitialize]}

\newcommand{\Trans}{{\mathrm{T}}}
\newcommand{\Herm}{{\mathrm{H}}}
\newcommand{\trace}{{\mathrm{tr}}}

\usepackage{titlesec}
\let\subparagraph\relax
\titlespacing{\section}{0pt}{6pt plus 2pt minus 1pt}{4pt plus 1pt minus 1pt} 
\titlespacing{\subsection}{0pt}{4pt plus 2pt minus 1pt}{2pt plus 1pt minus 1pt} 
\title{Enhanced Sparse Bayesian Learning Methods with Application to Massive MIMO Channel Estimation \vspace{-1mm}}
\author{
\IEEEauthorblockN{Arttu Arjas and Italo~Atzeni}
\IEEEauthorblockA{Centre for Wireless Communications, University of Oulu, Finland \\
E-mail: \{arttu.arjas, italo.atzeni\}@oulu.fi}
\thanks{This work was supported by the Research Council of Finland (336449 Profi6, 346208 6G~Flagship, and 348396 HIGH-6G).}\vspace{-2mm}}

\begin{document}

\maketitle

\begin{abstract}
We consider the problem of sparse channel estimation in massive multiple-input multiple-output systems. In this context, we propose an enhanced version of the sparse Bayesian learning (SBL) framework, referred to as \textit{enhanced SBL (E-SBL)}, which is based on a reparameterization of the original SBL model. Specifically, we introduce a scale vector that brings extra flexibility to the model, which is estimated along with the other unknowns. Moreover, we introduce a variant of E-SBL, referred to as \textit{modified E-SBL (M-E-SBL)}, which is based on a computationally more efficient parameter estimation. We compare the proposed E-SBL and M-E-SBL with the baseline SBL and with a method based on variational message passing (VMP) in terms of computational complexity and performance. Numerical results show that the proposed E-SBL and M-E-SBL outperform the baseline SBL and VMP in terms of mean squared error of the channel estimation in all the considered scenarios. Furthermore, we show that M-E-SBL produces results comparable with E-SBL with considerably cheaper computations.
\end{abstract}

\section{Introduction and Motivation} \label{sec:IM}    

Channel estimation is crucial to enable the beamforming design in coherent multiple-input multiple-output (MIMO) communications~\cite{Bjo17}. Accurate channel state information becomes even more important in the context of massive MIMO, where large antenna arrays allow for transmit/receive beamforming with extreme spatial resolutions. Channel sparsity, e.g., in the angular domain under far-field propagation, arises when few scatterers/reflectors are located in between the transmitter and receiver. Another example of sparse channel occurs when a user equipment (UE) is surrounded by scatterers and the signal seen by the base station (BS) is confined within a limited angle spread, as in the so-called one-ring channel model~\cite{Yin13,Pad19}. In general, when raising the operating frequencies to seek larger bandwidths~\cite{Raj20}, wireless channels tend to become sparse due to the increased spreading loss, penetration loss, and roughness of the materials relative to the wavelength.

Channel sparsity can be used to improve the channel estimation accuracy, e.g., by using compressive sensing algorithms~\cite{berger2010application}. Sparse Bayesian learning (SBL), initially proposed in~\cite{tipping2001sparse}, was utilized in~\cite{prasad2014joint,aminu2018bayesian,mishra2017sparse,wu2022clustered,srivastava2021sparse,salari2020joint} for sparse channel estimation. A related method utilizing variational message passing (VMP) was proposed for the same task in~\cite{pedersen2012application}. By adopting SBL, the channel is assumed to be Gaussian distributed after a suitable transformation, with each channel entry having its own variance. These variances can be thought of as weights specifying the significance of the channel entries and are iteratively estimated after marginalizing the channel out of the model.

In this work, we propose two enhanced SBL methods with application to massive MIMO channel estimation. The first method, referred to as \textit{enhanced SBL (E-SBL)}, is based on a reparameterization of the original SBL model. Additional flexibility is obtained by introducing individual scale parameters for the weights that are optimized along with the other unknowns, whereas the procedure used for the parameter estimation is analogous to that of SBL. The second method, referred to as \textit{modified E-SBL (M-E-SBL)}, is based on a computationally more efficient parameter estimation. The distributional assumptions for the channels and the corresponding variances are the same as in E-SBL. However, instead of marginalizing the channel out of the model, we target the joint posterior of the channels and weights. The posterior density can be partitioned in a way that simplifies the computation at each iteration, which leads to a more efficient iterative algorithm compared with E-SBL. This idea was first introduced in~\cite{calvetti2007gaussian}, where a hierarchical Bayesian model capable of recovering piecewise smooth signals was proposed. Similar sparsity-promoting models were recently studied in~\cite{senchukova2024bayesian} in the context of deconvolution, with Markov chain Monte Carlo methods used for the parameter estimation. In this paper, we study the differences of the proposed E-SBL and M-E-SBL with respect to the baseline SBL and VMP in terms of computational complexity and performance in the context of massive MIMO channel estimation. Note that, although the specific problem of massive MIMO channel estimation is considered in this paper, the proposed E-SBL and M-E-SBL can be regarded as general sparse estimation methods with a wide range of applications.

\section{System Model and Problem Formulation} \label{sec:SM}

Consider a massive MIMO system where $K$ single-antenna UEs communicate with an $M$-antenna BS in time-division duplexing mode. Let $\mathbf{H} = [\mathbf{h}_1, \dots, \mathbf{h}_K] \in \Compl^{M \times K}$ denote the uplink channel matrix, where $\mathbf{h}_k \in \Compl^{M}$ is the channel corresponding to UE~$k$. 
The channel matrix is generally unknown and can be estimated by means of uplink pilots. In this context, each UE simultaneously transmits a pilot of $N$ symbols to the BS through the unknown channel. The resulting signal observed at the BS is
\begin{align} \label{eq:Z}
    \mathbf{Z} = \mathbf{HP} + \mathbf{E} \in \mathbb{C}^{M \times N},
\end{align}
where $\mathbf{P} \in \mathbb{C}^{K \times N}$ is the pilot matrix collecting the pilots of all the UEs and $\mathbf{E}$ is a matrix of additive white Gaussian noise (AWGN) with independent $\mathcal{CN} (0, \sigma^2)$ entries.

In this work, we assume that the channel is sparse in some known domain. Hence, we can express the channel of UE~$k$ as $\mathbf{h}_k = \mathbf{Fu}_k \in \Compl^{M}$, where $\mathbf{F} \in \Compl^{M \times Q}$ denotes a known transform matrix, e.g., the discrete Fourier transform (DFT) matrix, and $\mathbf{u}_k \in \Compl^{Q}$ is a sparse vector (characterized by a considerable amount of approximately zero entries) to be estimated. We aim at exploiting the channel sparsity to enhance the channel estimation accuracy or, alternatively, to achieve the same accuracy with shorter pilots. With this formulation, the channel matrix can be expressed as $\mathbf{H} = \mathbf{FU}$,
where $\mathbf{U} \in \Compl^{Q \times K}$ is the transformed channel matrix that we now aim to estimate. To simplify the discussion in the following, we vectorize \eqref{eq:Z} as
\begin{align}
    \mathbf{z} = \mathrm{vec}(\mathbf{Z}) = \mathbf{A} \mathbf{u} + \mathbf{e} \in \Compl^{M N},
\end{align}
with $\mathbf{A} = \mathbf{P}^\Trans \otimes \mathbf{F} \in \Compl^{M N \times Q K}$, $\mathbf{u} = \mathrm{vec}(\mathbf{U}) = [u_{1}, \ldots, u_{QK}]^{\Trans} \in \Compl^{Q K}$, and $\mathbf{e} = \mathrm{vec}(\mathbf{E}) \in \Compl^{M N}$.

\subsection{SBL for MIMO Channel Estimation} \label{sec:SBL}

The MIMO channel estimation problem was tackled using SBL in, e.g., \cite{prasad2014joint}. In this section, we introduce the basic concept of SBL, on top of which we build our proposed E-SBL and M-E-SBL methods in Section~\ref{sec:Enhanced_SBL}. SBL is based on the Bayesian paradigm where one assigns prior distributions to the unknown parameters of a statistical model. In SBL, the parameters are assigned independent Gaussian priors with individual weights whose magnitude enforces the sparsity. The weights are inverse-gamma distributed and are estimated along with the main unknown $\u$. Specifically, introducing the vector of weights $\mathbf{w} = [w_1, \dots, w_{QK}]^\Trans \in \mathbb{R}_+^{QK}$, we assume
\begin{equation}
    \begin{split}
        u_j|w_j &\sim \mathcal{CN}(0, w_j), \\
        w_j &\sim \mathcal{IG}(\alpha, \beta).
    \end{split}    
\end{equation}
By setting $\alpha$ and $\beta$ to zero, we obtain uniform priors for the weights (in logarithmic scale), which are used in the baseline version of SBL. The aim is then to estimate the unknowns $\mathbf{u}$ and $\mathbf{w}$. This is done by marginalizing over $\mathbf{u}$ and maximizing the marginal likelihood
\begin{equation} \label{eq:SBL_obj}
    p(\mathbf{z}|\mathbf{w}) \propto f_\textrm{SBL}(\mathbf{w}) = \frac{1}{\det(\mathbf{V})}\exp \{-\mathbf{z}^\Herm\mathbf{V}^{-1}\mathbf{z}\}
\end{equation}
with respect to $\mathbf{w}$, where we have defined $\mathbf{V} = \mathbf{AWA}^\Herm + \sigma^2 \mathbf{I}_{MN} \in \Compl^{MN \times MN}$ and $\mathbf{W} = \mathrm{Diag}(\w) \in \Real_{+}^{QK \times QK}$. As there is no closed-form solution for this optimization problem, the weights are updated in an iterative manner using the expectation maximization (EM) algorithm (see \cite{tipping2001sparse} for details). The update formula for the weights at iteration $i$ is given by
\begin{equation} \label{eq:SBL_update}
    w_j^{i+1} = |\mu_j^i|^2 + \Sigma_{jj}^i, \quad \forall j = 1, \dots, QK,
\end{equation}
where $\mu_j^i$ and $\Sigma_{jj}^i$ are the $j$th entry and diagonal entry of
\begin{equation} \label{eq:SBL_update2}
    \begin{split}
        \boldsymbol{\mu}^i &= \frac{1}{\sigma^2}\boldsymbol{\Sigma}^i\mathbf{A}^\Herm \mathbf{z} \in \Compl^{QK}, \\
        \boldsymbol{\Sigma}^i &= \left(\frac{1}{\sigma^2} \mathbf{A}^\Herm \mathbf{A} + (\mathbf{W}^i)^{-1} \right)^{-1} \in \Compl^{QK \times QK},
    \end{split}
\end{equation}
respectively. It can be shown that $\mathbf{u}|\mathbf{z},\mathbf{w}^i \sim \mathcal{CN}(\boldsymbol{\mu}^i, \boldsymbol{\Sigma}^i)$. Thus, after convergence, $\boldsymbol{\mu}^i$ is used as an estimator for $\mathbf{u}$.

\section{Enhanced SBL for MIMO Channel Estimation} \label{sec:Enhanced_SBL}

In this section, we introduce an enhanced version of SBL, referred to as \textit{enhanced SBL (E-SBL)}, that allows more control over the sparsity level, which leads to a better performance. Moreover, we introduce a variant of E-SBL, referred to as \textit{modified E-SBL (M-E-SBL)}, with lower computational complexity. The reduced complexity is achieved by a different formulation of the objective function, which also leads to different performances in different setups.

\subsection{Enhanced SBL (E-SBL)}

While SBL is efficient in finding sparse solutions, it is not flexible in adapting to situations with different sparsity levels. To this end, we propose an enhanced version of SBL that is able to adapt to the sparsity level in a given situation. To achieve this, we introduce the scale vector $\boldsymbol{\tau} = [\tau_{1}, \ldots, \tau_{QK}]^{\Trans} \in \mathbb{R}_+^{QK}$ and the number of degrees of freedom $\nu \in \mathbb{R}_+$, and write
\begin{equation} \label{eq:E-SBL_model}
    \begin{split}
        u_j|w_j, \tau_j &\sim \mathcal{CN}(0, \tau_j w_j), \\
        w_j &\sim \mathcal{IG}(\nu/2, \nu/2), \\
        \tau_j &\sim \mathcal{IG}(\theta, \phi).
    \end{split}    
\end{equation}
This is equivalent to assuming that the $j$th entry of $\mathbf{u}$ follows the $t$ distribution with $\nu$ degrees of freedom and scale parameter $\tau_j$. Reducing $\nu$ makes the tails of the distribution heavier, so $\nu$ acts as a switch to control the sparsity level. The hyperparameters $\theta$ and $\phi$ also need to be prespecified, and tuning them will be discussed in Section~\ref{sec:NR}.

As in the baseline SBL described in Section~\ref{sec:SBL}, to estimate the model parameters, we marginalize over $\mathbf{u}$ and use Bayes' formula to find the marginal posterior density of $(\mathbf{w}, \boldsymbol{\tau})$ as
\begin{equation} \label{eq:E-SBL_obj}
    \begin{split}
        & p(\mathbf{w},\boldsymbol{\tau}|\mathbf{z}) \propto p(\mathbf{z}|\mathbf{w}, \boldsymbol{\tau}) p(\mathbf{w}) p(\boldsymbol{\tau}) = f_\textrm{E-SBL}(\mathbf{w}, \boldsymbol{\tau})\\
        &= \frac{1}{\det(\widetilde{\mathbf{V}})}\exp \{-\mathbf{z}^\Herm \widetilde{\mathbf{V}}^{-1}\mathbf{z}\} \prod_{j=1}^{QK} w_j^{-\frac{\nu + 2}{2}} \exp \left\{-\frac{\nu}{2 w_j}\right\} \\
        & \phantom{=} ~\times \prod_{j=1}^{QK} \tau_j^{-\theta - 1} \exp \left\{ -\frac{\phi}{\tau_j} \right\},
    \end{split}
\end{equation}
with $\widetilde{\mathbf{V}} = \mathbf{A}\widetilde{\mathbf{W}}\mathbf{A}^\Herm + \sigma^2 \mathbf{I}_{MN} \in \Compl^{MN \times MN}$ and $\widetilde{\mathbf{W}} = \mathrm{Diag}( \boldsymbol{\tau} \odot \mathbf{w}) \in \Real_{+}^{QK \times QK}$, where $\odot$ denotes element-wise multiplication. Then, the EM algorithm is again used to find $\mathbf{w}$ and $\boldsymbol{\tau}$ that maximize this density. The update formulas for the weights and scale parameters at iteration $i$ are given by
\begin{equation} \label{eq:E-SBL_update}
    \begin{split}
        w_j^{i+1} &= \frac{\nu/2 + (|\widetilde{\mu}_j^i|^2 + \widetilde{\Sigma}^i_{jj})/\tau_j^i}{\nu/2 + 2}, \quad \forall j = 1, \dots, QK,\\
        \tau_j^{i+1} &= \frac{\phi + (|\widetilde{\mu}_j^i|^2 + \widetilde{\Sigma}_{jj}^i)/w_j^{i+1}}{\theta + 2}, \quad \forall j = 1, \dots, QK,\\
    \end{split}
\end{equation}
where $\widetilde{\mu}_j^i$ and $\widetilde{\Sigma}_{jj}^i$ are the $j$th entry and diagonal entry of \vspace{-0.5mm}
\begin{equation}
    \begin{split}
        \widetilde{\boldsymbol{\mu}}^i &= \frac{1}{\sigma^2}\widetilde{\boldsymbol{\Sigma}}^i\mathbf{A}^\Herm \mathbf{z} \in \Compl^{QK}, \\
        \widetilde{\boldsymbol{\Sigma}}^i &= \left(\frac{1}{\sigma^2} \mathbf{A}^\Herm \mathbf{A} + (\widetilde{\mathbf{W}}^i)^{-1} \right)^{-1}  \in \Compl^{QK \times QK},
    \end{split}
\end{equation}
respectively (cf. \eqref{eq:SBL_update2}). As in the baseline SBL, after convergence, $\widetilde{\boldsymbol{\mu}}^i$ is used as an estimator for $\mathbf{u}$.

\subsection{Modified E-SBL (M-E-SBL)}

As can be seen from the update formulas \eqref{eq:SBL_update} and \eqref{eq:E-SBL_update} of the baseline SBL and E-SBL, respectively, the diagonal entries of the $QK \times QK$ matrices $\boldsymbol{\Sigma}$ and $\widetilde{\boldsymbol{\Sigma}}$ are needed to update the weights. For instance, the $j$th diagonal entry of $\boldsymbol{\Sigma}$ can be written as $\Sigma_{jj} = \e_j^\Trans \S^{-1} \e_j$, with $\S = \frac{1}{\sigma^2} \A^\Herm \A + \W \in \mathbb{C}^{QK \times QK}$ and where $\mathbf{e}_j \in \mathbb{R}^{QK}$ is a vector with one in the $j$th entry and zeros elsewhere. Computationally, this may be expensive depending on the values of $Q$ and $K$ and on the structure of $\mathbf{S}$. This motivates the M-E-SBL algorithm, which completely bypasses these computations. We start with exactly the same model parameterization \eqref{eq:E-SBL_model} as in E-SBL and the difference is in how the parameters are estimated. Instead of marginalizing over $\mathbf{u}$, we aim to maximize the joint posterior density of $\mathbf{u}$, $\mathbf{w}$, and $\boldsymbol{\tau}$. By Bayes' formula, the posterior density is \vspace{-1mm}
\begin{equation}
    \begin{split}
        & p(\mathbf{u}, \mathbf{w}, \boldsymbol{\tau}|\mathbf{z}) \propto p(\mathbf{z}|\mathbf{u})p(\mathbf{u}|\mathbf{w}, \boldsymbol{\tau})p(\mathbf{w})p(\boldsymbol{\tau}) \\
        &= f_\textrm{M-E-SBL}(\mathbf{u}, \mathbf{w}, \boldsymbol{\tau}) \\
        &= \exp\left\{-\frac{1}{\sigma^2}\|\mathbf{z} - \mathbf{Au}\|^2\right\} \frac{1}{\det(\widetilde{\mathbf{W}})}\exp \{-\mathbf{u}^\Herm \widetilde{\mathbf{W}}^{-1} \mathbf{u}\} \\
        & \phantom{=} ~\times \prod_{j=1}^{QK} w_j^{-\frac{\nu + 2}{2}} \exp \left\{-\frac{\nu}{2 w_j}\right\} \prod_{j=1}^{QK} \tau_j^{-\theta - 1} \exp \left\{ -\frac{\phi}{\tau_j} \right\}.
    \end{split}    
\end{equation}
As for \eqref{eq:SBL_obj} and \eqref{eq:E-SBL_obj}, there is no closed-form solution for this optimization problem. However, the objective function can be separated into three parts respective to $\mathbf{u}$, $\mathbf{w}$, and $\boldsymbol{\tau}$, and each part can be maximized separately with closed-form solution. Alternating among these solutions iteratively leads to a sequence that converges to a local maximizer of the objective. This is because the objective in each variable, while the other two are fixed, is a log-concave minorizer of the objective. Therefore, the value of the objective must increase at each iteration. The part containing $\mathbf{u}$ is given by \vspace{-1mm}
\begin{equation}
    \begin{split}
        f_\textrm{M-E-SBL}(\mathbf{u}|\mathbf{w}, \boldsymbol{\tau}) &= \exp\left\{-\frac{1}{\sigma^2}\|\mathbf{z} - \mathbf{Au}\|^2\right\} \\
        & \phantom{=} ~\times \frac{1}{\det(\widetilde{\mathbf{W}})}\exp\{-\mathbf{u}^\Herm \widetilde{\mathbf{W}}^{-1} \mathbf{u}\}.
    \end{split}
\end{equation}
Differentiating and setting the derivative to zero gives the update formula for $\mathbf{u}$ at iteration $i$: \vspace{-1mm}
\begin{equation}
    \mathbf{u}^{i+1} = \frac{1}{\sigma^2} \bigg( \frac{1}{\sigma^2} \mathbf{A}^\Herm \mathbf{A} +  (\widetilde{\mathbf{W}}^i)^{-1} \bigg)^{-1}\mathbf{A}^\Herm \mathbf{z}.
\end{equation}
The part containing $\mathbf{w}$ is given by \vspace{-1mm}
\begin{equation}
    \begin{split}
        f_\textrm{M-E-SBL}(\mathbf{w}|\mathbf{u}, \boldsymbol{\tau}) &= \frac{1}{\det(\widetilde{\mathbf{W}})}\exp \{-\mathbf{u}^\Herm \widetilde{\mathbf{W}}^{-1} \mathbf{u}\} \\
        & \phantom{=} ~\times \prod_{j=1}^{QK} w_j^{-\frac{\nu + 2}{2}} \exp \left\{-\frac{\nu}{2 w_j}\right\}.
    \end{split}
\end{equation}
Differentiating and setting the derivative to zero gives the update formula for $\mathbf{w}$ at iteration $i$: \vspace{-1mm}
\begin{equation} \label{eq:M-SBL_update_w}
    w_j^{i+1} = \frac{\nu/2 + |u^{i+1}_j|^2/\tau_j^i}{\nu/2 + 2}, \quad \forall j = 1, \dots, QK.
\end{equation}
Lastly, the part containing $\boldsymbol{\tau}$ is given by
\begin{equation}
    \begin{split}
        f_\textrm{M-E-SBL}(\boldsymbol{\tau}|\mathbf{u}, \mathbf{w}) &= \frac{1}{\det(\widetilde{\mathbf{W}})}\exp \{-\mathbf{u}^\Herm \widetilde{\mathbf{W}}^{-1} \mathbf{u}\} \\
        & \phantom{=} ~\times \prod_{j=1}^{QK} \tau_j^{-\theta - 1} \exp \left\{ -\frac{\phi}{\tau_j} \right\}. 
    \end{split}    
\end{equation}
Differentiating and setting the derivative to zero gives the update formula for $\boldsymbol{\tau}$ at iteration $i$:
\begin{equation} \label{eq:M-SBL_update_tau}
    \tau_j^{i+1} = \frac{\phi + |u^{i+1}_j|^2/w_j^{i+1}}{\theta + 2}, \quad \forall j = 1, \dots, QK.
\end{equation}

\begin{figure*}[!t]
    \centering
    \begin{subfigure}[b]{0.475\textwidth}
        \centering
        \hspace{-6mm} \input{figures/fig_3.tex} \vspace{-1mm}
        \caption{NMSE versus SNR ($M = 256$, $N = 12$, $3$ scatterers).}
        \label{fig:MSE_SNR}
    \end{subfigure}
    \hfill
    \begin{subfigure}[b]{0.475\textwidth}
        \centering
        \input{figures/fig_2.tex} \vspace{-1mm}
        \caption{NMSE versus pilot length ($\textrm{SNR} = 0$~dB, $M = 256$, $3$ scatterers).}   
        \label{fig:MSE_N}
    \end{subfigure}
    
    \vspace{5mm}
    
    \begin{subfigure}[b]{0.475\textwidth}
        \centering
        \input{figures/fig_1.tex} \vspace{-1mm}
        \caption{NMSE versus number of antennas ($\textrm{SNR} = 0$ dB, $N = 12$, $3$ scatterers).}   
        \label{fig:MSE_M}
    \end{subfigure}
    \hfill\begin{subfigure}[b]{0.475\textwidth}
        \centering
        \input{figures/fig_4.tex} \vspace{-1mm}
        \caption{NMSE versus number of scatterers ($\textrm{SNR} = 0$~dB, $M = 256$, $N = 12$).}   
        \label{fig:MSE_L}
    \end{subfigure}

    \vspace{5mm}
    
\caption{NMSE versus different parameters with $K=10$.}
\end{figure*}
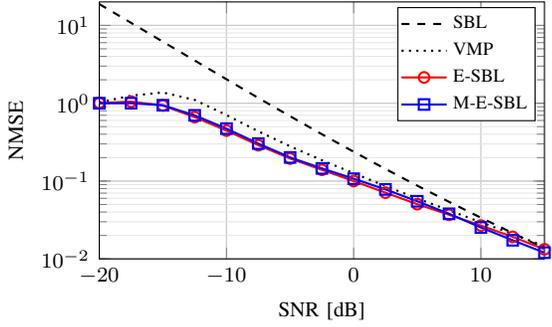
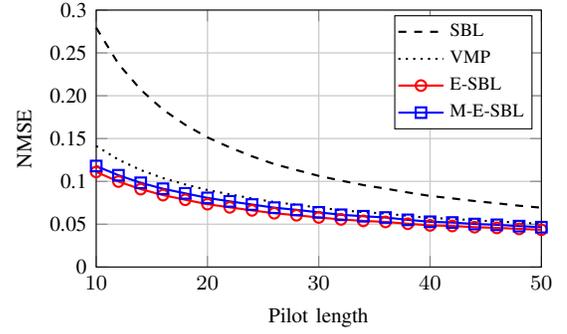
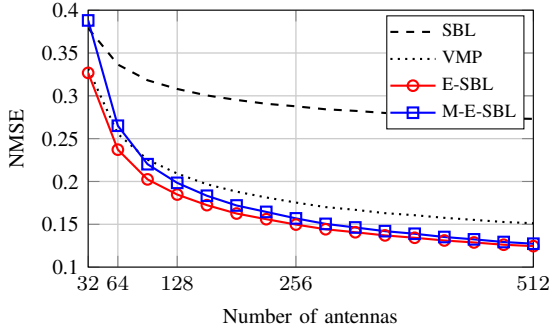
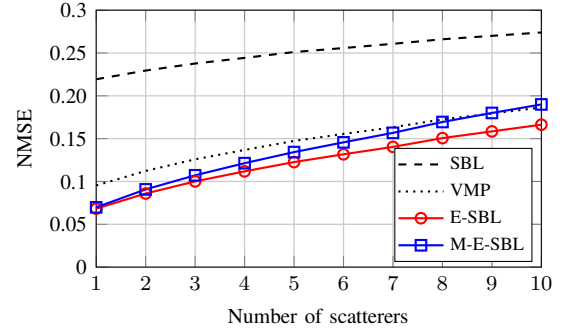

\subsection{Differences Between E-SBL and M-E-SBL} \label{sec:diff}

Although the model specification for E-SBL and M-E-SBL is the same, there is a substantial difference in how the objective functions of the two methods are defined. In E-SBL, the transformed channel $\mathbf{u}$ is marginalized over, leaving only the prior hyperparameters $\mathbf{w}$ and $\boldsymbol{\tau}$ to be estimated. Bayes' formula is used to form the marginal posterior density of the parameters (i.e., the objective function) and the EM algorithm is utilized to find a local optimizer. In M-E-SBL, Bayes' formula is used to form the joint posterior density of the transformed channel $\mathbf{u}$ and its prior hyperparameters $\mathbf{w}$ and $\boldsymbol{\tau}$ (i.e., the objective function). The model parameters are updated in an alternating iterative manner until a local optimizer of the objective is found. 
The effect of the marginalization can be seen by comparing the update formulas of $\mathbf{w}$ and $\boldsymbol{\tau}$ in E-SBL (i.e., \eqref{eq:E-SBL_update}) and M-E-SBL (i.e., \eqref{eq:M-SBL_update_w}, \eqref{eq:M-SBL_update_tau}). In E-SBL, the channel posterior covariance matrix $\widetilde{\mathbf{\Sigma}}$ is involved in the formulas, while in E-M-SBL it is not. This way, E-SBL takes the variability in the channel into account: this may be desirable, e.g., at very low signal-to-noise ratio (SNR), where the estimation uncertainty is high. However, there is an additional cost associated with the computation of the diagonal entries of $\widetilde{\mathbf{\Sigma}}$ at each iteration of the algorithm.

\subsection{Computational Complexity of E-SBL and M-E-SBL} \label{sec:compl}

Computationally, the most expensive part of both algorithms is solving a linear system involving the $QK \times QK$ matrix $\mathbf{S} = \frac{1}{\sigma^2} \mathbf{A}^\Herm \mathbf{A} + \widetilde{\mathbf{W}}^{-1}$, which in general has complexity $\mathcal{O}(Q^3K^3)$. In E-SBL, this has to be done $QK + 1$ times per iteration to find the diagonal values of $\mathbf{S}^{-1}$, while in M-E-SBL it has to be done only once per iteration. The complexity is reduced if the pilots and/or the transform matrix are orthogonal. This can be seen by using the properties of the Kronecker product, as
\begin{equation}
    \begin{split}
    \mathbf{A}^\Herm \mathbf{A} = (\mathbf{P}^\Trans \otimes \mathbf{F})^\Herm (\mathbf{P}^\Trans \otimes \mathbf{F}) = \mathbf{P} \mathbf{P}^\Herm \otimes \mathbf{F}^\Herm \mathbf{F}.
    \end{split}
\end{equation}
If both $\mathbf{P}^\Herm$ and $\mathbf{F}$ are orthogonal, $\mathbf{S}$ is diagonal and thus simple to invert. If only $\mathbf{P}^\Herm$ is orthogonal, $\mathbf{S}$ is block-diagonal and can be inverted block-wise. Lastly, if only $\mathbf{F}$ is orthogonal, $\mathbf{S}$ is a block matrix consisting of diagonal blocks. In these cases, the matrix inversion can be simplified by reordering the rows and columns.
Hence, depending on the case, M-E-SBL can provide a significant computational speed-up compared to E-SBL. Finally, we point out that the computational complexity of SBL is comparable to that of E-SBL, while the complexity of VMP is comparable to that of M-E-SBL.

\section{Numerical Results} \label{sec:NR}

We test E-SBL and M-E-SBL via numerical experiments and compare them with the baseline SBL \cite{tipping2001sparse} and VMP \cite{pedersen2012application}. As a metric, we use the normalized mean squared error (NMSE), defined as $\textrm{NMSE} = \frac{\mathbb{E}[\|\widehat{\mathbf{H}} - \mathbf{H}\|^2_\textrm{F}]}{\mathbb{E}[\|\mathbf{H}\|^2_\textrm{F}]}$, where $\widehat{\mathbf{H}}$ and $\mathbf{H}$ denote the estimated and true channel matrices, respectively. The expectations are computed by averaging over 1000 independent channel realizations. The channels are generated using a far-field channel model with a carrier frequency of $30$~GHz and assuming few scatterers in between the BS and the UEs. This gives rise to angular sparsity, as the signal seen by the BS comes from few directions corresponding to the scatterers. The locations of the scatterers are drawn from a uniform distribution between $100$ and $500$~meters from the BS with an angular spread of $\frac{\pi}{6}$ radians, and the angular deviations are also drawn from a uniform distribution (independently for each UE). The channel matrix is normalized at the end to maintain the desired SNR. We test the algorithms' performance varying the SNR, pilot length, number of antennas, and number of scatterers. For the pilot matrix $\mathbf{P}$, we use the first $K$ rows of an $N \times N$ DFT matrix and, for the transformation matrix $\mathbf{F}$, we use an $M \times M$ DFT matrix. For both the E-SBL and M-E-SBL, we set the number of degrees of freedom to $\nu = 1$, and $\theta = \phi = 10^{-2}$ to give an uninformative prior for $\boldsymbol{\tau}$.


Fig.~\ref{fig:MSE_SNR} plots the NMSE of the channel estimation against the SNR. E-SBL and M-E-SBL yield comparable NMSEs and outperform both SBL and VMP, especially at low SNR. The NMSEs of all the methods converge to about the same value when the SNR increases. Fig.~\ref{fig:MSE_N} shows the NMSE against the pilot length. Again, E-SBL and M-E-SBL perform clearly better than SBL and slightly better than VMP, with a larger gap for shorter pilot lengths. Fig.~\ref{fig:MSE_M} plots the NMSE against the number of antennas. With few antennas, E-SBL and VMP perform similarly and slightly better than both M-E-SBL and SBL. When the number of antennas increases, M-E-SBL surpasses VMP at around $M = 100$. For $M > 128$, both proposed methods perform clearly better than SBL and slightly better than VMP. Lastly, Fig.~\ref{fig:MSE_L} shows the NMSE against the number of scatterers. When the number of scatterers increases, the NMSE increases for all the methods due to reduced sparsity. With few scatterers, E-SBL and M-E-SBL perform better than both SBL and VMP. When the number of scatterers exceeds $9$, VMP starts to perform better than M-E-SBL, likely because M-E-SBL over-enforces sparsity. In the numerical results, the average number of iterations needed to reach convergence is $35$ for SBL, $34$ for VMP, $29$ for E-SBL, and $19$ for M-E-SBL.

\section{Conclusions} \label{sec:conclusions}

In this work, we developed two enhanced SBL methods, E-SBL and M-E-SBL. The proposed algorithms utilize a reparameterization of the original SBL model, which endows them with extra flexibility. The baseline SBL and E-SBL have similar computational complexity, while M-E-SBL is less complex. We showed through numerical results that the proposed methods outperform the baseline SBL and VMP in the context of sparse channel estimation in several MIMO scenarios. Future work will consider sub-THz communications enabled by hybrid analog-digital beamforming architectures and characterized by near-field channel propagation.

\addcontentsline{toc}{chapter}{References}
\bibliographystyle{IEEEtran}
\bibliography{refs_abbr,refs}

\end{document}

%% file: Definitions.tex
\newcommand{\e}{\mathbf{e}}

\renewcommand{\u}{\mathbf{u}}

\newcommand{\w}{\mathbf{w}}

\newcommand{\A}{\mathbf{A}}

\renewcommand{\S}{\mathbf{S}}

\newcommand{\W}{\mathbf{W}}

\newcommand{\Real}{\mbox{$\mathbb{R}$}}
\newcommand{\Compl}{\mbox{$\mathbb{C}$}}



%% file: figures/fig_3.tex
\begin{tikzpicture}

\begin{axis}[
	width=7.5cm,
	height=5cm,
	xmin=-20, xmax=15,
	ymin=10^-2, ymax=2*10^1,
	xlabel={SNR [dB]},
	ylabel={NMSE},
	xlabel near ticks,
	ylabel near ticks,
	x label style={font=\footnotesize},
	y label style={font=\footnotesize},
	ticklabel style={font=\footnotesize},
	legend style={at={(0.98,0.98)}, anchor=north east},
	legend style={font=\scriptsize, inner sep=1pt, fill opacity=0.75, draw opacity=1, text opacity=1},
	legend cell align=left,
	grid=both,
	major grid style={line width=0.5pt, draw=gray!40},
	grid style={line width=.2pt, draw= gray!20},
	ymode=log,
	title={},
	title style={font=\scriptsize, yshift=-2mm},
]

\addplot[thick, black, dashed]
table [x=Var1, y=err_SBL_tot, col sep=comma] {figures/files_txt/MSE_SNR_2.txt};
\addlegendentry{SBL};

\addplot[thick, black, dotted]
table [x=Var1, y=err_VMP_tot, col sep=comma] {figures/files_txt/MSE_SNR_2.txt};
\addlegendentry{VMP};

\addplot[thick, red, mark=o]
table [x=Var1, y=err_E_SBL_tot, col sep=comma] {figures/files_txt/MSE_SNR_2.txt};
\addlegendentry{E-SBL};

\addplot[thick, blue, mark=square]
table [x=Var1, y=err_M_E_SBL_tot, col sep=comma] {figures/files_txt/MSE_SNR_2.txt};
\addlegendentry{M-E-SBL};

\end{axis}

\end{tikzpicture}

%% file: figures/fig_2.tex
\begin{tikzpicture}

\begin{axis}[
	width=7.5cm,
	height=5cm,
	xmin=10, xmax=50,
	ymin=0, ymax=0.3,
	xlabel={Pilot length},
	ylabel={NMSE},
	ytick={0,0.05,0.1,0.15,0.20,0.25,0.3},
	yticklabels={0,0.05,0.1,0.15,0.20,0.25,0.3},
	xlabel near ticks,
	ylabel near ticks,
	x label style={font=\footnotesize},
	y label style={font=\footnotesize},
	ticklabel style={font=\footnotesize},
	legend style={at={(0.98,0.98)}, anchor=north east},
	legend style={font=\scriptsize, inner sep=1pt, fill opacity=0.75, draw opacity=1, text opacity=1},
	legend cell align=left,
	grid=both,
	major grid style={line width=0.5pt, draw=gray!40},
	grid style={line width=.2pt, draw=gray!20},
	title={},
	title style={font=\scriptsize, yshift=-2mm},
]

\addplot[thick, black, dashed]
table [x=Var1, y=err_SBL_tot, col sep=comma] {figures/files_txt/MSE_N_2.txt};
\addlegendentry{SBL};

\addplot[thick, black, dotted]
table [x=Var1, y=err_VMP_tot, col sep=comma] {figures/files_txt/MSE_N_2.txt};
\addlegendentry{VMP};

\addplot[thick, red, mark=o]
table [x=Var1, y=err_E_SBL_tot, col sep=comma] {figures/files_txt/MSE_N_2.txt};
\addlegendentry{E-SBL};

\addplot[thick, blue, mark=square]
table [x=Var1, y=err_M_E_SBL_tot, col sep=comma] {figures/files_txt/MSE_N_2.txt};
\addlegendentry{M-E-SBL};

\end{axis}

\end{tikzpicture}

%% file: figures/fig_1.tex
\begin{tikzpicture}

\begin{axis}[
	width=7.5cm,
	height=5cm,
	xmin=32, xmax=512,
	ymin=0.05, ymax=0.35,
	xlabel={Number of antennas},
	ylabel={NMSE},
	xtick={32,64,128,256,512},
	ytick distance=0.05,
	yticklabels={0.05,0.1,0.15,0.2,0.25,0.3,0.35,0.4},
	xlabel near ticks,
	ylabel near ticks,
	x label style={font=\footnotesize},
	y label style={font=\footnotesize},
	ticklabel style={font=\footnotesize},
	legend style={at={(0.98,0.98)}, anchor=north east},
	legend style={font=\scriptsize, inner sep=1pt, fill opacity=0.75, draw opacity=1, text opacity=1},
	legend cell align=left,
	grid=both,
	major grid style={line width=0.5pt, draw=gray!40},
    grid style={line width=.2pt, draw=gray!20},
	title={},
	title style={font=\scriptsize, yshift=-2mm},
]

\addplot[thick, black, dashed]
table [x=Var1, y=err_SBL_tot, col sep=comma] {figures/files_txt/MSE_M_2.txt};
\addlegendentry{SBL};

\addplot[thick, black, dotted]
table [x=Var1, y=err_VMP_tot, col sep=comma] {figures/files_txt/MSE_M_2.txt};
\addlegendentry{VMP};

\addplot[thick, red, mark=o]
table [x=Var1, y=err_E_SBL_tot, col sep=comma] {figures/files_txt/MSE_M_2.txt};
\addlegendentry{E-SBL};

\addplot[thick, blue, mark=square]
table [x=Var1, y=err_M_E_SBL_tot, col sep=comma] {figures/files_txt/MSE_M_2.txt};
\addlegendentry{M-E-SBL};

\end{axis}

\end{tikzpicture}

%% file: figures/fig_4.tex
\begin{tikzpicture}

\begin{axis}[
	width=7.5cm,
	height=5cm,
	xmin=1, xmax=10,
	ymin=0, ymax=0.3,
	xlabel={Number of scatterers},
	ylabel={NMSE},
	xtick={1,2,3,4,5,6,7,8,9,10},
	ytick={0,0.05,0.1,0.15,0.20,0.25,0.3},
	yticklabels={0,0.05,0.1,0.15,0.20,0.25,0.3},
	xlabel near ticks,
	ylabel near ticks,
	x label style={font=\footnotesize},
	y label style={font=\footnotesize},
	ticklabel style={font=\footnotesize},
	legend style={at={(0.98,0.02)}, anchor=south east},
	legend style={font=\scriptsize, inner sep=1pt, fill opacity=0.75, draw opacity=1, text opacity=1},
	legend cell align=left,
	grid=both,
	major grid style={line width=0.5pt, draw=gray!40},
	grid style={line width=.2pt, draw= gray!20},
	title={},
	title style={font=\scriptsize, yshift=-2mm},
]

\addplot[thick, black, dashed]
table [x=Var1, y=err_SBL_tot, col sep=comma] {figures/files_txt/MSE_L_2.txt};
\addlegendentry{SBL};

\addplot[thick, black, dotted]
table [x=Var1, y=err_VMP_tot, col sep=comma] {figures/files_txt/MSE_L_2.txt};
\addlegendentry{VMP};

\addplot[thick, red, mark=o]
table [x=Var1, y=err_E_SBL_tot, col sep=comma] {figures/files_txt/MSE_L_2.txt};
\addlegendentry{E-SBL};

\addplot[thick, blue, mark=square]
table [x=Var1, y=err_M_E_SBL_tot, col sep=comma] {figures/files_txt/MSE_L_2.txt};
\addlegendentry{M-E-SBL};

\end{axis}

\end{tikzpicture}